\documentclass[sigconf]{acmart}

\renewcommand\footnotetextcopyrightpermission[1]{} 
\usepackage{subfig}
\usepackage{algorithm}
\usepackage{algpseudocode}
\usepackage{amsmath}

\setcopyright{rightsretained}

\begin{document}


\title{A Frequency-Based Learning-To-Rank Approach for Personal Digital Traces}

\author{Daniela Vianna}
\affiliation{%
  \institution{Dept. of Computer Science}
  \institution{Rutgers University}
  \city{New Jersey}
  \state{USA}
}
\email{dvianna@cs.rutgers.edu}

\author{Am\'elie Marian}
\affiliation{%
  \institution{Dept. of Computer Science}
  \institution{Rutgers University}
  \city{New Jersey}
  \state{USA}
}
\email{amelie@cs.rutgers.edu}
\renewcommand{\shortauthors}{D. Vianna et al.}


\begin{abstract}
Personal digital traces are constantly produced by connected devices, internet services and interactions. These digital traces are typically small, heterogeneous and stored in various locations in the cloud or on local devices, making it a challenge for users to interact with and search their own data. By adopting a multidimensional data model based on the six natural questions --- what, when, where, who, why and how --- to represent and unify heterogeneous personal digital traces, we can propose a learning-to-rank approach using the state of the art LambdaMART algorithm and frequency-based features that leverage the correlation between content (what), users (who), time (when), location (where) and data source (how) to improve the accuracy of search results. Due to the lack of publicly available personal training data, a combination of known-item query generation techniques and an unsupervised ranking model (field-based BM25) is used to build our own training sets. Experiments performed over a publicly available email collection and a personal digital data trace collection from a real user show that the frequency-based learning approach improves search accuracy when compared with traditional search tools.
\end{abstract}

\maketitle

%
%
\section{Introduction}

Digital traces of our lives are constantly being produced and saved by users, as personal files, emails, social media interactions, multimedia objects, calendar items, contacts, GPS tracking of mobile devices, records of financial transactions, etc. Digital traces are usually small, heterogeneous and stored in various locations in the cloud or on local devices making it hard for users to access and search their own data. 

Search of personal data is usually focused on retrieving information that users know exists in their own dataset, even though most of the time they do not know in which source or device they have seen the desired information. Personal search have been throughout studied in specific real-life scenarios as desktop search~\cite{stuff} and email search~\cite{halawi2015}. In this work, we tackle search over an integrated personal dataset comprised by a multitude of heterogeneous personal digital traces from a variety of data sources. In this scenario, each personal digital trace is a source of knowledge and can be related to different data traces by shared common information. The richness of contextual information attached to the digital data can be of great help to users searching for information they remember having stored and accessed in the past. This information can naturally be modeled following the six contextual dimensions: \emph{what}, \emph{who}, \emph{when}, \emph{where}, \emph{why} and \emph{how}. 

Learning-to-rank approaches have been very successful in solving real-world ranking problems. However, the existing models for ranking are trained on either explicit relevance judgments (crowdsourced or expert-labeled) or clickthrough logs. For personal digital traces, none of these is available nor pursuable, in addition, there is a dearth of synthetic personal datasets and benchmarks. To overcome those challenges, we propose a learning-to-rank approach that relies on a combination of known-item query generation techniques and an unsupervised ranking model (field-based BM25) to heuristically build training sets. Furthermore, since personal digital traces are results of actions and events of users, the correlation between objects can be leverage to improve the accuracy of search results. In this regard, we represent the input data by a set of frequency-based features that takes into consideration the correlation between content (\emph{what}), users (\emph{who}), time (\emph{when}), location (\emph{where}) and data source (\emph{how}). In this work, we use a state-of-the-art learning-to-rank algorithm based on gradient boosted decision trees, LambdaMART~\cite{Burges2010}, to learn a ranking model to map feature vectors to scores. The work presented in this paper is developed as part of a series of tools to let user retrieve, store and organize their digital traces {\em on their own devices} guaranteeing some clear privacy and security benefits.

In this paper, we make the following contributions:
\begin{itemize}
    \item A representative feature set to represent query-matching object pairs built upon a novel frequency-based feature space that leverages entities interactions within and across dimensions in the dataset,
    \item A novel combination of known-item query generation techniques and an unsupervised ranking model to heuristically generate labeled training sets,
    \item A quantitative evaluation of the proposed search technique, as well as comparison with two popular search methodologies: \emph{BM25} and \emph{field-based BM25}. Our results show that moderately large datasets can benefit from learning techniques when combined with a compact frequency-based feature set. 
\end{itemize}

%
%
\section{Data Model}
\label{datamodel}

Considering the fact that contextual cues are strong trigger for autobiographical memories (~\cite{brewer88, PIMbook, sevensinsmemory, Wagenaar86}), and that personal data is rich in contextual information, in the form of metadata, application data, or environment knowledge, personal digital traces can be represented following a combination of dimensions that naturally summarize various aspects of the data collection: \emph{who}, \emph{when}, \emph{where}, \emph{what}, \emph{why} and \emph{how}. In this work, we use an intuitive multidimensional data model that relies on these six dimensions as the unifying features of each personal digital trace object, regardless of its source~\cite{asist2019}. Below is a list of dimensional data that can be extracted from a user's personal digital traces:

\begin{itemize}
    \item what: content \\
Messages, messages subjects, description of events/users, publications, list of interests of a user
\item who: with whom, from whom, to whom \\
User names, senders, recipients, event owners, list of friends, authors
\item where: physical or logical, in the real-world and in the system \\
Hometown, location, event venues, URLs, file/folder paths
\item when: time and date, but also what was happening concurrently \\
Birthday, file/message/event created-/modified- time
\item why: sequences of data/events that are connected
\item how: application, author, environment
\end{itemize}

 \begin{figure}[ht!]
  \centering
  \includegraphics[scale=0.37]{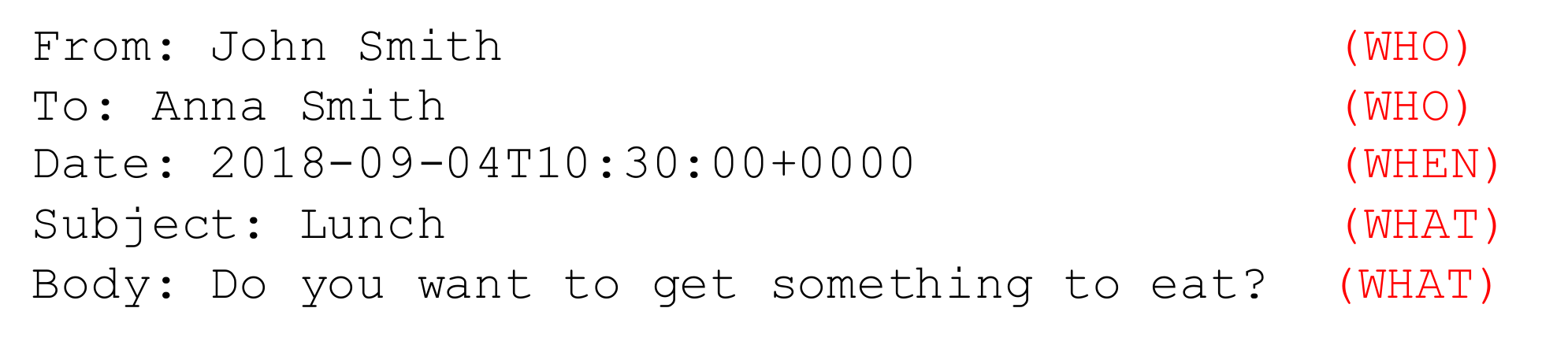}
  \caption[]{Simplified example of a user email message classified according to the 6 contextual dimensions model. }
  \label{fig:gmail_dim}
\end{figure}

Figure~\ref{fig:gmail_dim} shows a digital trace from a Gmail message with each piece classified as belonging to one of the six contextual dimensions: \emph{what, who, where, when, why and how}. By using the six proposed dimensions, multiple digital traces are unified and can be linked to each other, even though they could be from different sources and have their own data schema. For instance, the email message from Figure~\ref{fig:gmail_dim} can be linked to different data traces that have John Smith under the same dimension \emph{who}. Those traces could be, for example, a Facebook post tagging John Smith or a Calendar entry shared with John Smith.

Having defined the multidimensional data model, it is still necessary to find an effective mechanism to automatically translate the heterogeneous set of personal data into the six dimensions. To this end, we use a machine learning multi-class classifier using a combination of LSTM (Long Short-Term Memory) and Dense layers~\cite{asist2019}. Given a sentence, the classifier will output a label (\emph{who}, \emph{when}, \emph{where}, \emph{what}, \emph{why} and \emph{how}). For instance, the subject in an email is a sentence with label \emph{what}. During the classification process, the \emph{when} dimension is normalized allowing dates from different sources to be matched. Normalization also makes possible partial matches. For instance, a query searching for June, will match objects with time June 2016 and June 2017 regardless of the original format of the data. Entity resolution is applied to the parsed \emph{who} and \emph{where} dimensions identifying separate instances of the same entity in data traces coming from the same sources, and across sources. We use the classifier to translate raw data retrieved from third-party sources into the 6 dimension model, without the need of human intervention.

%
%
\section{Scoring Model}

Unlike Web search, where the focus is often on discovering new relevant information, search in personal datasets is typically focused on retrieving data that the user knows exists in their dataset. Besides, users have unique habits and interpretations of their own data. In this scenario, standard search techniques are not ideal as they do not leverage the additional knowledge the user is likely to have about the target object, or the connections between objects pertaining to a given user. This extra knowledge, represented by the six dimensional model introduced in Section~\ref{datamodel}, can be leveraged to provide rich and accurate search capabilities over personal digital traces.

As personal digital traces are very specific to each user and are constantly evolving over time, it is necessary to find a scoring model that can generalize well over user-specific datasets. Learning-to-rank approaches have proved to be very efficient to solve ranking problems. However, learning algorithms require a huge amount of data to generalize well given the large feature space usually employed (from thousands to tens of thousands of features), due to the curse of dimensionality~\cite{bellman}. To be able to employ a learning approach on available personal datasets, which are typically not very large, we adopted a compact feature set based on a representation of the input data using the six contextual dimensions presented in Section~\ref{datamodel}. 

In Section~\ref{sec:methodology}, we define matching objects and queries in our personal search scenario. The 34 features proposed are detailed in Section~\ref{w5hfeatures}. In Section~\ref{sec:what} we continue to discuss the frequency score focusing on the \emph{what} dimension that is comprised mostly of text. Section~\ref{l2rmodel} describes the learning-to-rank algorithm used in this work to generate our ranking model and validate the feature space proposed.

\subsection{Scoring Methodology}
\label{sec:methodology}

In this work, we consider each digital trace from a personal dataset to be a distinct object that can be returned as the result to a query. 

\begin{definition}[Object in the Integrated Dataset]
\label{def:object}
An object $O$ in the dataset is a structure that has fields corresponding to the 6 dimensions mentioned earlier. Each of these dimensions contains 0 or more items (corresponding to text, entities identified by entity resolution, times, locations, etc).
\end{definition}

Formal queries have the same structure as objects in the unified dataset.

\begin{definition}[Query]
\label{def:query}
A query $Q$ over the dataset is represented as an object as defined in Definition~\ref{def:object}.
\end{definition}

Given objects \textit{Q} and \textit{O}, \textit{O} is considered as an answer to object \textit{Q} treated as a query if it contains at least one dimension/item specified in \textit{Q}. Unfortunately users' memories are notoriously unreliable~\cite{sevensinsmemory, croft}, and fully trusting their recollection of contextual information can lead to miss relevant results. In looking for (partially) matching objects to a given query, each dimension will be searched separately, and the results will be combined generating a list of candidates with some partial order. With the intention of finding an optimal order for this list of candidates, in the following sections we introduce our learning-to-rank approach on top of a representative feature set built from our novel frequency-based feature space.

\subsection{Frequency-based Features}
\label{w5hfeatures}

Because personal digital traces are byproducts of actions and events of users, they are not independent objects. Our intuition is that the correlation between traces (objects) can be leveraged to improve
the accuracy of search results. With that in mind, we explore how the correlations between users (\emph{who}), time (\emph{when}), location (\emph{where}), topics (\emph{what}) and data sources (\emph{how}) can be used to improve search over personal data. We exploit those interactions and correlations by way of a frequency score. For each dimension and combination of dimensions we compute a score that will be used later as features to represent the input data in our learning-to-rank approach.

Frequencies can be computed for individual users or group of users. They can be associated with multiple times, multiple data sources, and also with a set of locations. For example, from a set of emails exchanged between a group of users, we can extract the frequency (number of interactions) with which those users communicated, and in which time period those interactions occurred. In short, frequency expresses the strength of relationships, based on users, time, location, content and data sources (\emph{who, when, where, what, how}). To keep it simple, every time an interaction occurs, the frequency is increased by one; however, in practice, the frequency function allows us to weigh differently distinct types of interactions. For example, likes or comments on a Facebook post could be weighed differently, giving more relevance to interactions coming from comments than likes. Different roles, e.g. From and To in an email, can also be weighed differently. 

To take advantage of the strong correlation between group of users, which is an important feature of personal corpora, we also compute the frequency between group of users, source, times and location. Other application scenarios could also benefit from our six-dimension model, with other groups highlighted in a dedicated frequency-based scoring. In this work, we use a set of 34 features to represent the input data. The feature set is comprised by 30 features resulting from all possible combinations between the dimensions \emph{who, what, when, where} and \emph{how} plus 4 extra features that model the correlation between group of users (\emph{who groups}); group of users and time (\emph{who groups, when}); group of users and data source (\emph{who groups, how}); and finally, group of users, time and location (\emph{who groups, when, where}). The feature vector is defined in Definition~\ref{def:featurevector}.

\begin{definition}[Feature Vector]
\label{def:featurevector}
\textbf{x} = $[x_{1} \ldots x_{34}]$ is a feature vector comprised by 34 frequency-based features. Each feature $x_{i}$ is computed by a frequency function $f(S_{i},Q,O)$, where $S_{i} \in \mathcal{S}$. $\mathcal{S}$ represents all possible combinations between the 5 dimensions \emph{who, what, when, where} and \emph{how}, plus the 4 extra features that model the correlation between group of users. $Q$ is a query (Definition~\ref{def:query}) and $O$ is an object in the user dataset (Definition~\ref{def:object}).
\end{definition}

To illustrate our query and scoring methodology consider the following search scenario: the user is interested in a message from 2018 (\emph{when}), sent by John (\emph{who}), about the topic ``Lunch'' (\emph{what}). We can define query $Q_{1}$ as (\emph{when}: 2018,  \emph{who}: John, \emph{what}: Lunch). By Definition~\ref{def:object}, the object in Figure~\ref{fig:gmail_dim} ($O_{1}$) is a matching to the given query ($Q_{i}$) containing all dimension/item specified in the query -- \emph{when}:2018, \emph{who}:John, and \emph{what}:``Lunch''. The query-object pair ($Q_{1}, O_{1}$) can be represented by a 34 frequency-based feature vector $\textbf{x} = [x_{1} \ldots x_{34}]$ as introduced in Definition~\ref{def:featurevector}. Each feature $x_{i}$ represents the frequency score for a set of dimensions $S_{i}$, query $Q_{i}$ and object $O_{i}$:

\begin{itemize}
  \item[] $x_{1} = f((\text{what:Lunch}),Q_{1},O_{1})$ 
  \item[] $x_{2} = f((\text{who:John}),Q_{1},O_{1})$
  \item[] $x_{3} = f((\text{when:2018}),Q_{1},O_{1})$
  \item[] $x_{6} = f((\text{what:Lunch, who:John}),Q_{1},O_{1})$
  \item[] $x_{7} = f((\text{what:Lunch, when:2018}),Q_{1},O_{1})$
  \item[] $x_{9} = f((\text{what:Lunch, how:Gmail}),Q_{1},O_{1})$
  \item[] $x_{10} = f((\text{who:John, when:2018}),Q_{1},O_{1})$
  \item[] $x_{12} = f((\text{who:John, how:Gmail}),Q_{1},O_{1})$ 
  \item[] $x_{16} = f((\text{what:Lunch,who:John, when:2018}),Q_{1},O_{1})$ 
  \item[] $x_{18} = f((\text{what:Lunch,who:John, how:Gmail}),Q_{1},O_{1})$ 
  \item[] $x_{20} = f((\text{what:Lunch,when:2018, how:Gmail}),Q_{1},O_{1})$ 
  \item[] $x_{23} = f((\text{who:John, when:2018, how:Gmail}),Q_{1},O_{1})$ 
  \item[] $x_{27} = f((\text{what:Lunch,who:John,when:2018,how:Gmail}), Q_{1},O_{1})$
\end{itemize}

If a set of dimensions $S_{i}$ is not present in query $Q_{1}$ and object $O_{i}$, the frequency score $f(S_{i},Q,O) = 0$. When a query $Q$ contain multiple values for a dimension e.g., Q = (who: {John, Alice}), the values for each feature $x_{i}$ will be the summation of the frequencies for each individual value.

To understand how frequencies ($f(S_{i},Q,O)$) are computed, consider the following example: lets assume a dataset $D$ containing 10 objects that mention John under the \emph{who} dimension, being 4 of those 10 objects from Facebook and the remaining 6 from Gmail. Given object $O_{1}$ and query $Q_{1}$ from the previous example, we can say that the frequency of John ((who:John)) in dataset $D$ for query $Q_{1}$ and matching object $O_{1}$ is $x_{2} = f((\text{who:John}),Q_{1},O_{1})$, where $f((\text{who:John}),Q_{1},O_{1}) = 10$. We can also say that the frequency of John in Gmail ((who:John,how:Gmail)) in dataset $D$ for query $Q_{1}$ and matching object $O_{1}$ is $x_{12} = f((\text{who:John, how:Gmail}),Q_{1},O_{1})$, where $f((\text{who:John, how:Gmail}),Q_{1},O_{1}) = 6$.

Notice that the \emph{why} dimension is not explored in this paper, but is the topic of related work~\cite{odbase,cikmdemo}. This dimension can be derived by inference and could be used to connect different fragments of data that derive from a common real-life task, or episode. 

\subsection{Scoring the \emph{What} Dimension}
\label{sec:what}

The \emph{what} dimension in the six-dimension model is composed of content information comprising mostly of text. Based on that fact, we use two standard text approaches to link and score objects for the \emph{what} dimension: field-based BM25 (used to score the what dimension alone) and topic modeling~\cite{steyvers_griffiths07} (used to link the what dimension with the other dimensions). 

\textbf{Field-based BM25.} A field-based BM25 is a state-of-the-art TF-IDF type of ranking function that takes into consideration the document structure. In our scenario, the fields in the field-based BM25 correspond to the 5 dimensions proposed, \emph{what, who, when, where} and \emph{how}. To compute the field-based BM25 score for the \emph{what} dimension, we use a popular full-text search platform from the Apache Lucene project, Solr~\cite{ApacheSolr}. All data retrieved for a user is unified and parsed according with the six dimensions (Section~\ref{datamodel}) and then, exported to Solr. For each user query, we search Solr using the values from the \emph{what} dimension, getting as a result a partial list of matching documents with its respective field-based BM25 score. Even though Solr contains the data for all 5 dimensions, we are only interested in use field-based BM25 to score the \emph{what} dimension, since this dimension contains most of the content of an object. For the remaining dimensions, we use our frequency-based function as introduced in Section~\ref{w5hfeatures}.

\textbf{Topic Modeling.} A ``Topic'' consists of a cluster of words that frequently occur together. Topic models use contextual cues to find connections between words with similar meanings and to distinguish between use of words with multiple meanings. Given a document, we would like to identify what possible topics have generated that data. In our case, topic modeling would be an important feature to connect different objects, including objects from different data sources. The association between topics (\emph{what}), user (\emph{who}), times (\emph{when}), location (\emph{where}) and source (\emph{where}) could shed some light on finding objects that could be a better matching to the user query. To define topics for each object in the user data set, we use a topic model package called MALLET and a text collection built from the content classified under the \emph{what} dimension for each object in the user data set. The MALLET~\cite{McCallumMALLET} topic model package includes a fast and scalable implementation of Gibbs sampling. The Gibbs Sampling algorithm considers each word token in the text collection in turn, and estimates the probability of assign the current word token to each object, conditioned on the topic assignments to all other word tokens. For each object in the user data set, MALLET computes the topic composition of documents. We use the most relevant topic for each document to cluster documents per topic. For each document in a topic, we extract the person/entity mentioned in \emph{who} dimension, the times from \emph{when}, location from \emph{where} and source from \emph{how}. With that information, we are able to build the correlation between person/entity, times, location and source for each topic (\emph{what}). Also, we are able to estimate the frequency of those correlation/interactions using the frequency function presented in Section~\ref{w5hfeatures}. Besides the topic composition of documents, MALLET also outputs the words in the corpus with their topic assignments and frequencies. We use this list of words per topic and the words specified in the user query (for the \emph{what} dimension), to find the topic that are a more close representative of the user query. Then, we can use the topic that matches the query to find out a partial list of documents that are matching candidates to the query, based solely on the contents of the \emph{what} dimension.

\begin{figure}[ht!]
  \centering
   \includegraphics[scale=0.37]{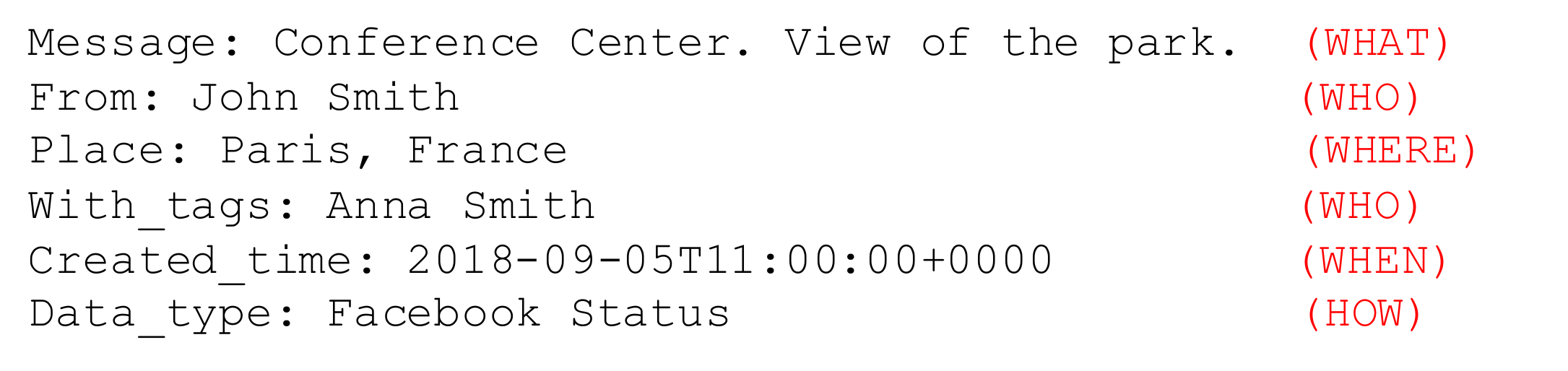}
  \caption[]{Simplified example of a user Facebook post classified according to the 6 contextual dimensions model. }
  \label{fig:facebook_dim}
\end{figure}

To illustrate how topic modeling can support our search, consider $T$ a topic composed by the following key words: hotel, lunch, street, trip, miles, view, lake, ride, restaurant and conference. Assuming that topic $T$ is the most relevant topic for object $O_{1}$ in Figure~\ref{fig:gmail_dim}, and object $O_{2}$ in Figure~\ref{fig:facebook_dim}, we can say that objects $O_{1}$ and $O_{2}$ are correlated by their \emph{what} dimension. By considering all objects (documents) clustered under the same topic $T$, we can learn how strong person/entity (\emph{who}), times (\emph{when}), location (\emph{where}) and source (\emph{how}) are connected with relation to a topic (\emph{what}). Again, this strength is measured by a way of a frequency score as presented in Section~\ref{w5hfeatures}. 

\subsection{Learning-to-Rank Model}
\label{l2rmodel}

In the previous sections, we explained how query-document pairs are represented by a feature vector built upon our frequency-based feature space. To map the feature vector to a real-valued score we need to train a ranking model. Our choice of learning-to-ranking algorithm is the state-of-the-art LambdaMART~\cite{Burges2010}. LambdaMART uses gradient boosted decision trees, which incrementally builds regression trees trying to correct the leftover error from the previous trees. At the end, the prediction model is an ensemble of weaker prediction models that complement each other for robustness. During a training phase, we must define the best set of parameters that results in a robust and accurate model. For this cross-validation stage, we will consider the following parameters:
\begin{itemize}
    \item tree: number of trees in the ensemble
    \item leaf: maximum number of leaves per tree
    \item mls: minimum number of samples each leaf has to contain
    \item shrinkage: learning rate
    \item metric: training metric to be optimized for
\end{itemize}

In the next section, we will discuss how training and evaluation sets can be built for personal data search, a scenario where publicly available personal training data does not exist.

\section{Query Sets}
\label{querysets}

In a learning-to-rank algorithm, each pair of query-document(object) is represented by a vector of numerical features. In addition to the feature vector, pairs of query-documents could be augmented with some relevance information. Then, a model has to be trained to map the feature vector to a score. One of the challenges of using learning-to-rank for personal data search is to be able to build a training set without human intervention or any external information (e.g., expert labeling or click data). To this end, in this section we present a combination of heuristics that given a user dataset is able to simulate a human-labeled training set to tailor the learning model to each specific user dataset.

Search of personal data is usually focused on retrieving information that users know exists in their own data set. Considering the fact that personal data search is a known-item type of search, simulated queries can be automatically generated, using known-item query~\cite{Elsweiler07towardstask-based} generation techniques such as the ones presented in~\cite{Azzopardi:2007} and~\cite{Kim:2009}. In this work, queries are created by randomly choosing a set of dimensions (\emph{who, what, when, where, how}) and values/items (e.g. email's Subject, Facebook post's content) from a target object, as described in Algorithm~\ref{KnownItemAlgorithm}. Each call to Algorithm~\ref{KnownItemAlgorithm} will result in a query-target object pair.

 \begin{algorithm} 
\caption{Known-item query generation algorithm.}
\label{KnownItemAlgorithm}
\begin{algorithmic}[1]
\Procedure{Build\textendash Query(dataset D)}{}
\State $Q = ()$ {\em\small /* Initialize query Q. */}
\State {\em\small /* Randomly choose a target object $O_i$ from the dataset D */}
\State $O_i =$ random(D) 
\State {\em\small /* Select dimensions */}
\State $d =$ select\_dimensions(\{what, who, when, where, how\})
\For {each $d_i \in d$}
\State {\em\small /* Randomly choose $v$ values from target object $O_i$ and dimension $d_i$ */}
\State $v$ = select\_values($d_i$)
\State {\em\small /* Add dimension and values to the query $Q$ */}
\State $Q(d_i) = v$
\EndFor
\State return $Q$
\EndProcedure
\end{algorithmic}
\end{algorithm}

By using the proposed known-item query generation technique, we are able to build a list of query-target object pairs. However, a learning-to-rank training set is composed not only by pairs of query-known document, but also by a list of matching documents per query. In~\cite{Dehghani2017}, the authors use classic unsupervised information retrieval models, such as BM25, as a weak supervision signal for training deep neural ranking models. In a similar fashion, we adopt an unsupervised ranking model, field-based BM25, to retrieve matching objects to a given query. In Section~\ref{sec:what}, we explained how the data retrieved for a user is unified and parsed according with the six dimensions (Section~\ref{datamodel}) and then, the parsed data is exported to Solr where it can be searched using a field-based BM25 approach. Given a query generated by Algorithm~\ref{KnownItemAlgorithm}, a call to Solr will retrieve a list of matching documents to this query --- the list is ranked using field-based BM25. Now, for each query, we have a list of matching documents that includes the (generated) target object and its corresponding feature vector as described in Section~\ref{w5hfeatures}. Since the target object is known for this query, a relevance label of $1$ is assigned to it; otherwise, the relevance label will be $0$.

%
%
\section{Case Studies 1: Personal Digital Data Traces} 

We evaluate the efficacy of the proposed learned ranking model by comparing its performance with two popular scoring methodologies: \emph{BM25} and \emph{field-based BM25}. Experiments are performed over real user data from a variety of data sources. 

\subsection{Methodology}

\subsubsection{\bf Data Set.}
\label{sec:dataset}

\begin{table}[h]
\centering
\begin{tabular}{|l||r|r|}
\hline
\textbf{Data Source}     & \textbf{\#Objs}      & \textbf{Size}           \\ \hline \hline
Facebook     	  & 3875    & 28Mb        \\ \hline
Gmail           	  & 28318  & 3Gb           \\ \hline
Dropbox       	  & 573      & 32Mb        \\ \hline
Foursquare   	  & 55        & 59Kb         \\ \hline
Twitter         	  & 3929    & 22Mb         \\ \hline
Google Calendar & 330      & 620Kb        \\ \hline
Google+             & 110      & 367Kb        \\ \hline
Google Contacts & 525     & 629Kb         \\ \hline
Bank            & 412  &  415Kb      \\ \hline
Firefox         & 181921 & 63Mb  \\ \hline \hline
Total                  & 219,993 & 3.6Gb          \\ \hline
\end{tabular}
\caption{Personal dataset.}
\label{table:data-statistics}
\vspace{-0.25in}
\end{table}

There is a dearth of synthetic data sets and benchmarks to evaluate search over personal data. Thus, we perform our evaluation using a real dataset collected by a personal data extraction tool~\cite{DataExtraction} containing approximately two hundred thousand objects. 

Table~\ref{table:data-statistics} shows the composition of our real user dataset, including the number and size of objects retrieved from different sources over different periods of time. The dataset was automatically classified according with the 6 contextual dimensions: \emph{what, who, when, where, why} and \emph{how} (Section~\ref{datamodel}). We used this unified dataset to evaluate the frequency-based learning-to-rank approach proposed.

\subsubsection{\bf Training and Evaluation query sets.}
\label{sec:query_set}

We train and evaluate our model on the user's own device using heuristically generated samples. As detailed in Section~\ref{querysets}, each query is automatically created by randomly choosing a target object from the evaluation data set. We then choose \emph{d} dimensions, from which we randomly select $v$ random values. For this set of experiments, we built a training set comprised by 19000 queries over our personal dataset (Table~\ref{table:data-statistics}). To built the query sets, we use $v=1$ and 4 different values for parameter $d$: \{\emph{what, who}\}, \{\emph{what, who, when}\}, \{\emph{what, who, when, how}\}, and \{\emph{what, who, how}\}. The evaluation set was built in a similar fashion. Approximately 6000 queries were heuristically generated using the same combination of parameters as the training set. Since less than 2\% of objects in the user dataset have location, the dimension \emph{where} was not included in the query sets.

\subsubsection{\bf Evaluation Techniques and Metrics.}
\label{subsec:evaltechniques}

We evaluate the efficacy of the proposed approach by comparing it with two popular scoring methodologies: \emph{BM25} and \emph{field-based BM25}. 

\textbf{BM25} is a state-of-the-art type of TF-IDF function that ranks a list of matching documents based on the query content that appears in each document. To be able to use BM25 with the retrieved dataset (Section~\ref{sec:dataset}), the heterogeneous and decentralized digital traces have to be integrated in one unified collection. It is done by exporting the data retrieved to a unified data collection in Solr~\cite{ApacheSolr}, a popular open source full-text search platform from the Apache Lucene project. This approach allows user to search for information across the entire set of retrieved digital traces, which is already a significant step forward from the current state, where users have to search each data source individually.

\textbf{Field-based BM25} is a version of BM25 that takes into consideration the structure of a document. In our scenario, the fields in the field-based
BM25 correspond to the five dimensions proposed: \emph{what, who, when, where, how}. As described in Section~\ref{sec:what}, before being exported to Solr, the retrieved dataset (Section~\ref{sec:dataset}) is unified and parsed according with the six dimensions (Section~\ref{datamodel}). It allows for the dataset to be searched using field-based BM25 with each field corresponding to a respective dimension. Note that by using the five dimensions, \textbf{we are giving the field-based BM25 approach the advantage of using our multidimensional data model to unify and organize the user data}.

The scoring model proposed is evaluated using 4 standard evaluation metrics: Mean Reciprocal Rank (MRR) of the top-ranked 50 documents, success (precision) of the top 1 retrieved document (success@1), success of the top 3 retrieved document (success@3), and success of the top 10 retrieved document (success@10). Wilcoxon signed-rank test with $p\_value < 0.05$ is used to determine statistically significant differences.

\begin{table}[h!]
\begin{tabular}{|l||rrrr|}
\hline 
Method           & MRR & s@1 & s@3 & s@10 \\ \hline \hline
BM25             & 0.3629 & 0.2701 & 0.4104 & 0.5350 \\ \hline
Field-based BM25 & 0.5082 & 0.4252 & 0.5502 & 0.6690 \\
w5h-l2r          & 0.5184 & 0.4406 & 0.5601 & 0.6900 \\
\hline
\end{tabular}
\caption{MRR, s@1 (success@1), s@3 (success@3), s@10 (success@10) for all $6000$ queries (groups $1$ to $4$) from the Personal Digital Data Traces dataset (Table~\ref{table:data-statistics}). Compared against the baseline (\emph{BM25}), the results are statistically significant (Wilcoxon signed-rank test).}
\label{table_all}
\end{table}

\subsubsection{\bf Ranking Model.} 
\label{subsec:rankingmodel}

To train and evaluate our model, we use the LambdaMART implementation provided by the RankLib library~\cite{ranklib}. RankLib is a library of learning to rank algorithms that is part of The Lemur Project~\cite{lemur}.

The first step in our evaluation was to define the best set of parameters that would give us a more robust and accurate model. The parameters evaluated are: number of trees (tree); number of leaves for each tree (leaf); minimum leaf support (mls), minimum number of samples each leaf has to contain; and, training/evaluation metric (metric). In our evaluation we considered the following parameters: 

\begin{itemize}
    \item tree: 50, 100, 250 and 500
    \item leaf: 10, 15, 35 and 45
    \item mls: 10, 20 and 50
    \item shrinkage: 0.01, 0.03, 0.1, 0.3, 0.5, 1.0
    \item metric: MRR (Mean Reciprocal Rank)
\end{itemize}

We adopted a 5-fold cross validation process to estimate the performance of each model. After the validation process, we selected the model that shows the best performance on the training set, also taking into account the spread between training and testing metrics. The model selected, that we will call \emph{w5h-l2r}, has the following parameters: number of trees $= 50$; number of leaves $= 15$; minimum leaf support $= 10$; shrinkage $= 0.1$.

\subsection{\bf Results.} 

In Table~\ref{table_all} we compare the ranking performance of the baseline (\emph{BM25}), \emph{field-based BM25} and learned ranking model (\emph{w5h-l2r}) with respect to the entire evaluation set composed by approximately 6000 queries heuristically generated as described in Section~\ref{sec:query_set}. The results show that both search models using the data parsed according to the multidimensional data model (Section~\ref{datamodel}), \emph{field-based BM25} and \emph{w5h-l2r}, outperform the keyword-based approach, \emph{BM25}, for MRR, success@1, success@3 and success@10. It shows that traditional keyword-based search methods are not appropriate in a setting where users may remember valuable contextual cues to guide the search. Observe that the learned ranking model, \emph{w5h-l2r}, outperform the \emph{field-based BM25} approach for all 4 evaluation metrics, \textbf{showing that moderately large datasets can also benefit from learning-to-rank techniques when paired with a representative feature set built from our novel frequency-based feature space}. 

We now conduct a more thorough evaluation by dividing the evaluation set (Section~\ref{sec:query_set}) in four different groups by the dimensions in each query as described in Table~\ref{groups_parms}.

 \begin{table}[h!]
 \centering
\begin{tabular}{|l||l|}
\hline
Groups  & Dimensions             \\ \hline \hline
Group 1 & what, who              \\
Group 2 & what, who, when        \\
Group 3 & what, who, when, how   \\
Group 4 & what, who, how  \\ 
\hline
\end{tabular}
\caption{Dimensions used to generate four groups of queries.}
\label{groups_parms}
\end{table}

\begin{table*}[h!]
\centering
\begin{tabular}{c}
\subfloat[Group 1: what, who]{
\begin{tabular}{|l||cccc|}
\hline
Method           & MRR & s@1 & s@3 & s@10 \\ \hline \hline
BM25             &  0.3435 &  0.2450  & 0.3939  & 0.5290  \\ \hline
Field-based BM25 &  0.4407 &  0.3605  & 0.4809  & 0.6000  \\
w5h-l2r          &  0.4432 &  0.3730  & 0.4851  & 0.6430  \\ \hline
\end{tabular}
} \\
\subfloat[Group 2: what, who, when]{
\begin{tabular}{|l||cccc|}
\hline
Method           & MRR & s@1 & s@3 & s@10 \\ \hline \hline
BM25             &  0.3759 & 0.2875  & 0.4224  & 0.5370 \\ \hline
Field-based BM25 &  0.5760 & 0.4850  & 0.6261  & 0.7420 \\
w5h-l2r          &  0.5970 & 0.5084  & 0.6465  & 0.7660 \\ \hline
\end{tabular}
} \\ 
\subfloat[Group 3: what, who, when, how]{
\begin{tabular}{|l||cccc|}
\hline
Method           & MRR & s@1 & s@3 & s@10 \\ \hline \hline
BM25             &  0.4213 & 0.3223  & 0.4614  & 0.5870 \\ \hline
Field-based BM25 &  0.6168 & 0.5215  & 0.6417  & 0.7810 \\
w5h-l2r          &  0.6331 & 0.5272  & 0.6618  & 0.7940 \\ \hline
\end{tabular}
} \\
\subfloat[Group 4: what, who, how]{
\begin{tabular}{|l||cccc|}
\hline
Method           & MRR & s@1 & s@3 & s@10 \\ \hline \hline
BM25             &  0.3484 & 0.2591  & 0.3888  & 0.5200 \\ \hline
Field-based BM25 &  0.4621 & 0.3863  & 0.4974  & 0.6150 \\
w5h-l2r          &  0.4632 & 0.3964  & 0.4902  & 0.6000 \\ \hline 
\end{tabular}
}   \\
\end{tabular}
\caption{MRR, s@1 (success@1), s@3 (success@3), s@10 (success@10) for groups $1$,$2$,$3$, and $4$ from the Personal Digital Data Traces dataset (Table~\ref{table:data-statistics}). Compared against the baseline (\emph{BM25}), the results are statistically significant (Wilcoxon signed-rank test).}
\label{table:groups}
\end{table*}

Table~\ref{table:groups}a-d, show the MRR, success@1, success@3 and success@10 of each search approach, \emph{BM25} (baseline), \emph{field-based BM25}, and \emph{w5h-l2r}, for Group 1 to 4 of queries. For all 4 groups, the search approaches that use the data classified according with our multidimensional data model are considerably more accurate than the keyword-based approach, \emph{BM25}, confirming the importance of including contextual information to improve search accuracy when searching personal data. When compared against each other, \emph{field-based BM25} and \emph{w5h-l2r}, the learned ranking model outperform the \emph{field-based BM25} model for all four groups; however, the improvements were more relevant for Group 2 (\emph{what, who, when}) and Group 3 (\emph{what, who, when, how}), showing that for this dataset, using the proposed learning model and training data, the \emph{when} dimension and all related features played an important role in scoring query-document pairs. The results for \emph{w5h-l2r} when compared with \emph{field-based BM25} are statistically significant (Wilcoxon signed-rank test, $p\_value < 0.05$) for Groups 2 and 3, evaluation metric MRR and success@k. For Group 1 the results are not statistically significant for MRR and success@3. For Group 4, the results are not statistically significant for MRR, success@1 and success@3.

\begin{figure*}[t!]
\centering
\begin{tabular}{ccc}
\subfloat[Group 1]{
    \centering
    \includegraphics[scale=0.5]{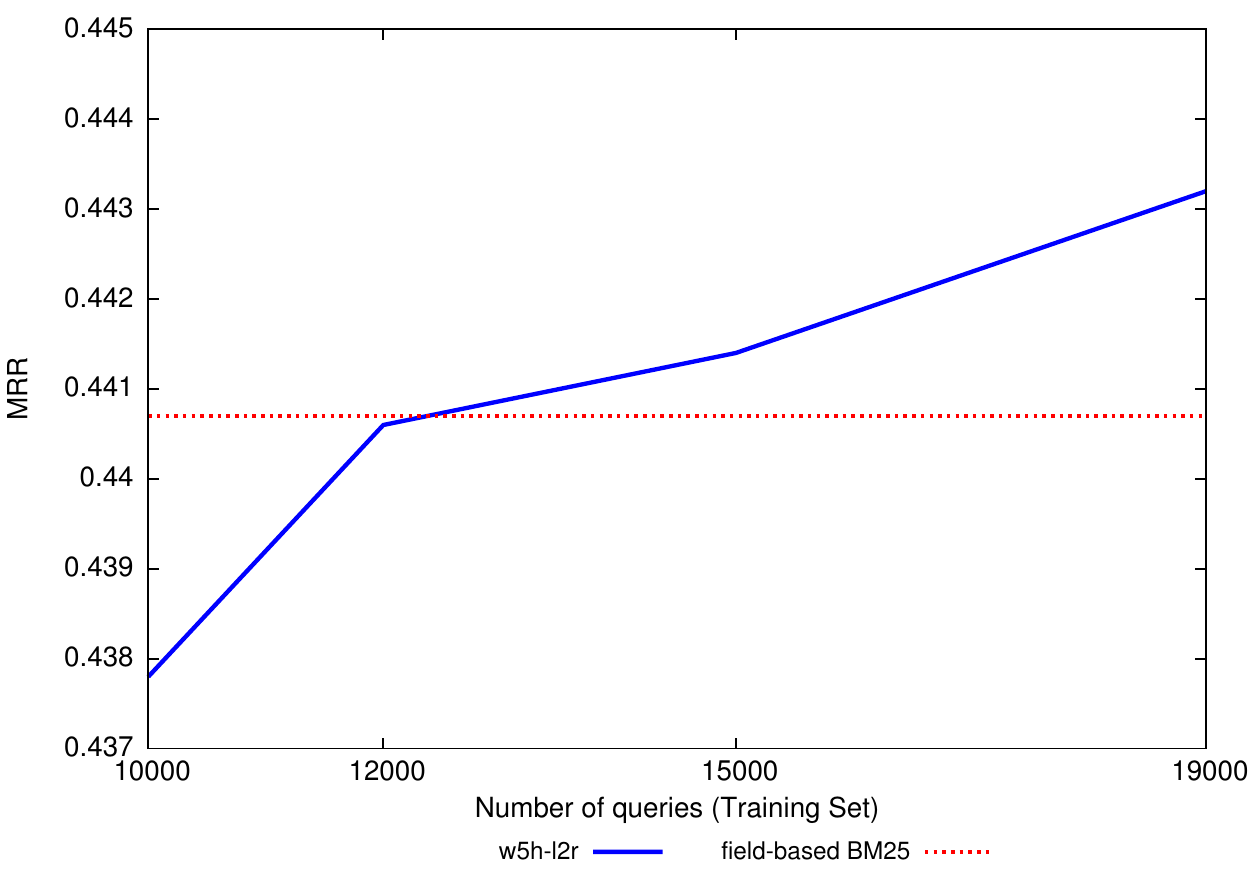} 
} &
\subfloat[Group 2]{
    \centering
        \includegraphics[scale=0.5]{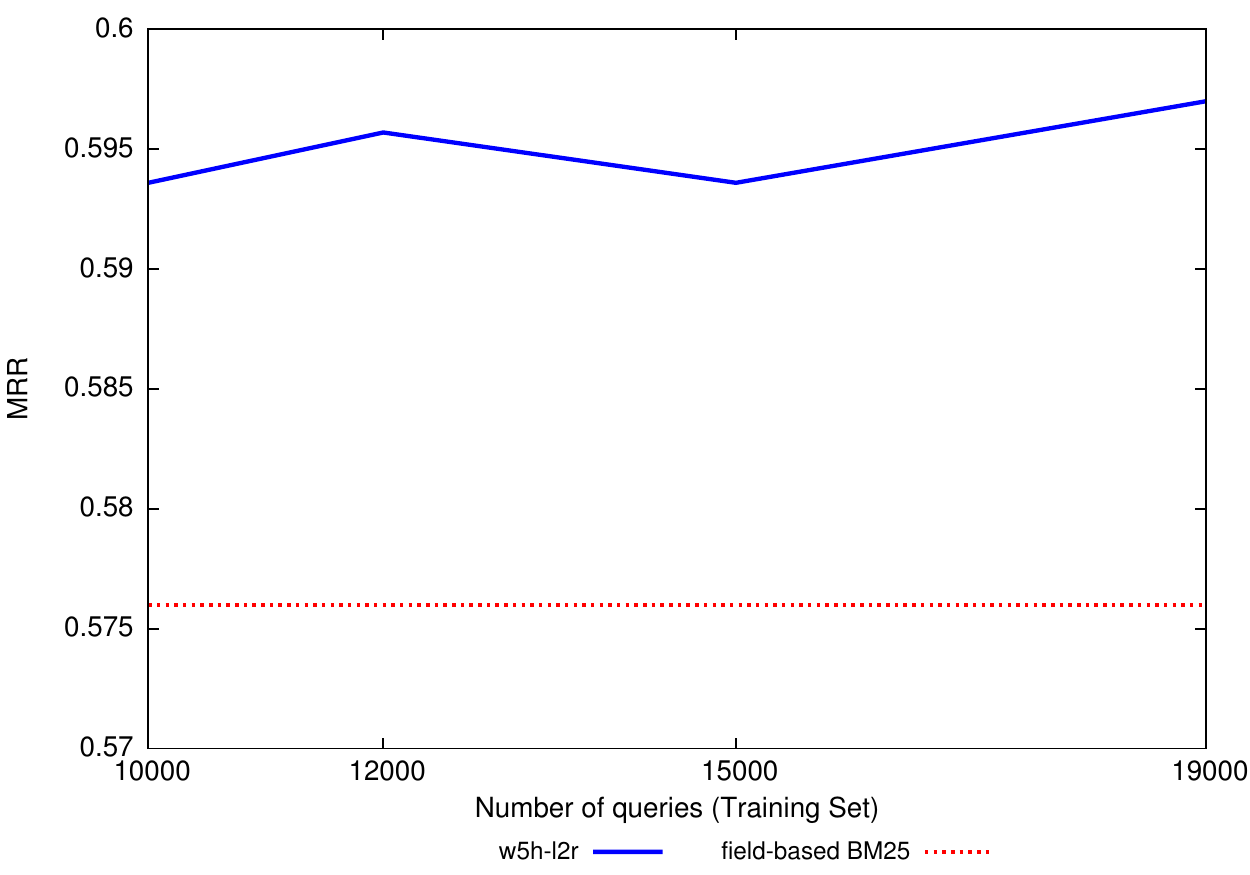}  
} \\ 
\subfloat[Group 3]{
    \centering
        \includegraphics[scale=0.5]{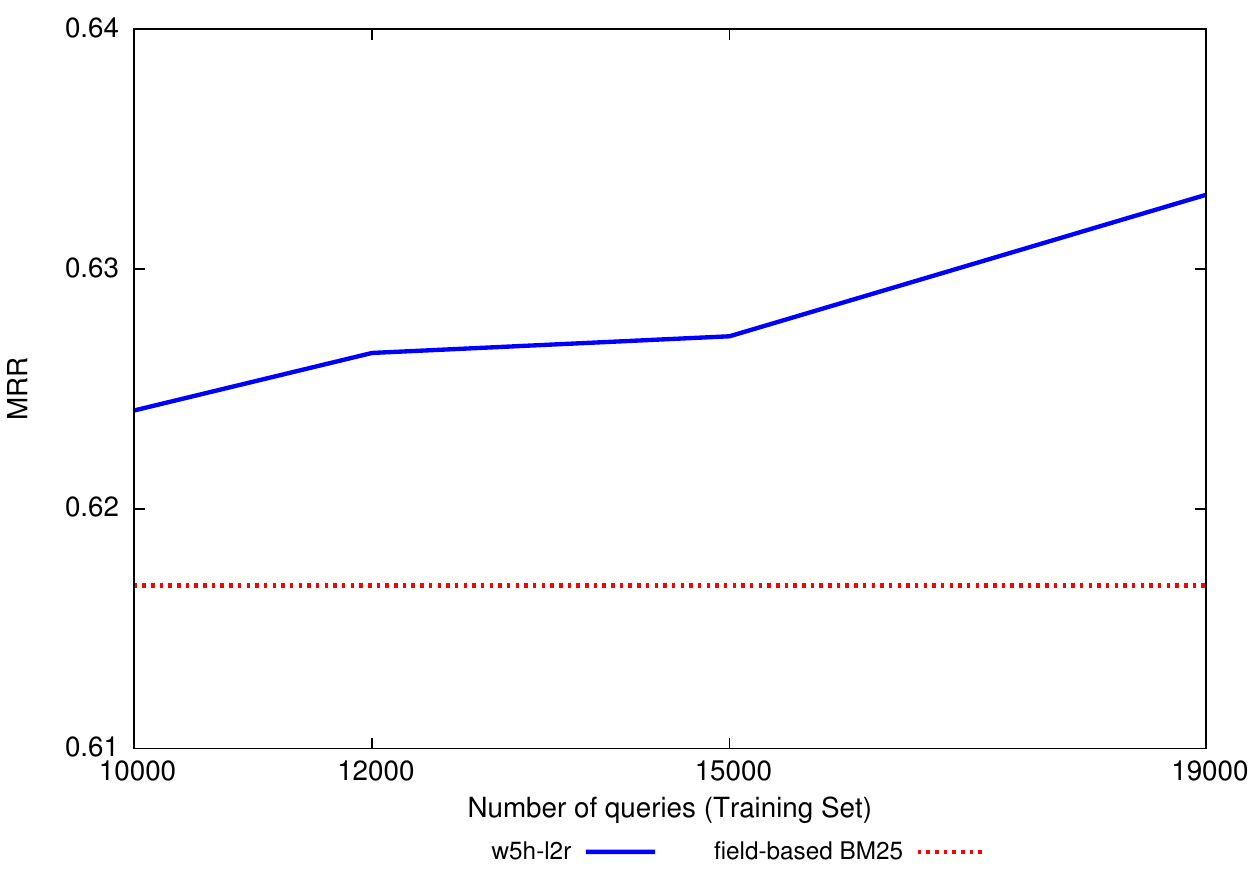} 
} &
\subfloat[Group 4]{
    \centering
        \includegraphics[scale=0.5]{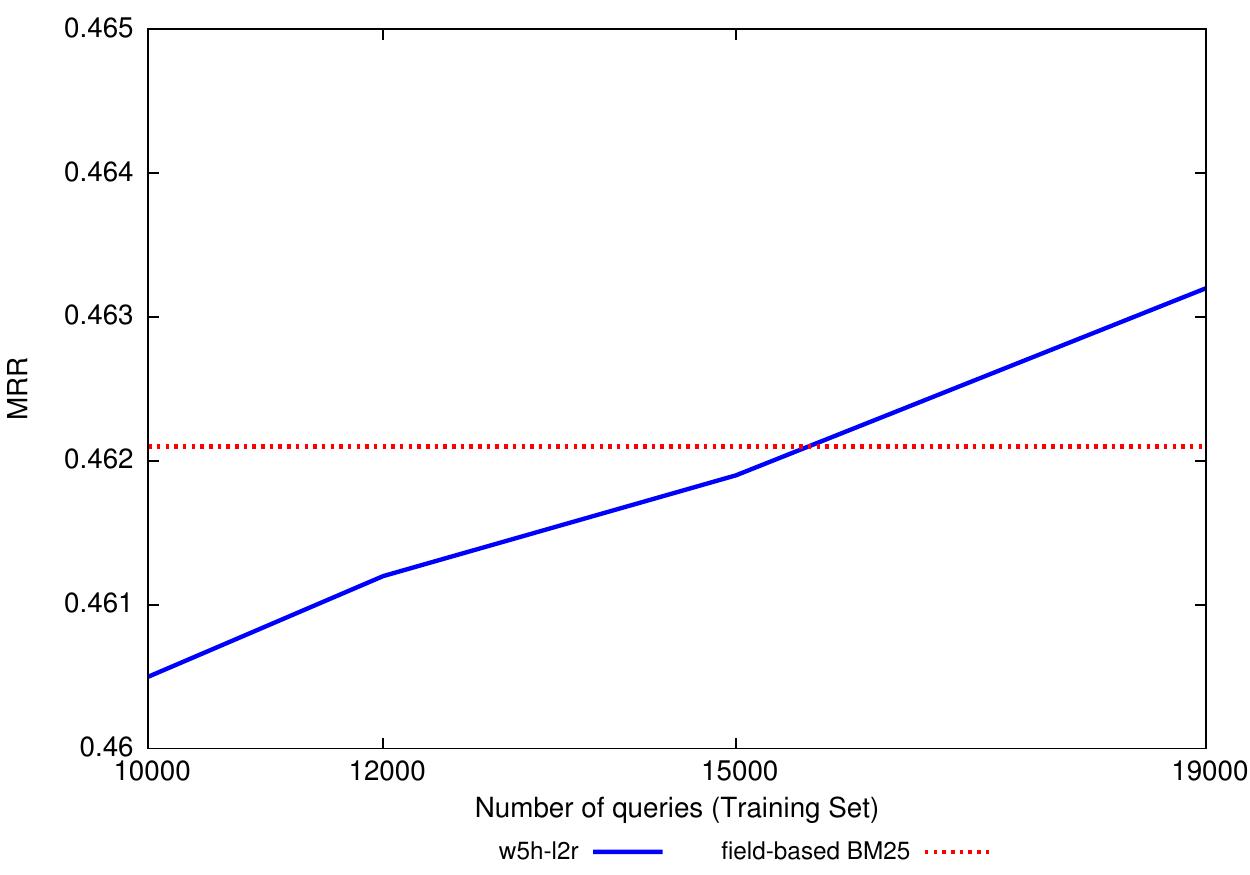} 
}  \\
\end{tabular}
\caption{Performance (MRR) of the learning model, \emph{w5h-l2r}, for groups 1, 2, 3 and 4 of queries from the Personal Digital Data Traces dataset (Table~\ref{table:data-statistics}) as the number of training samples increases.}
\label{fig:mapbytrainingsize}  
\end{figure*}

Figure~\ref{fig:mapbytrainingsize} presents the performance (MRR) of the learning model, \emph{w5h-l2r}, for Group 1 to 4 of queries as the number of training samples increases. We observe that the performance of the learned ranking model (\emph{w5h-l2r}) clearly improves as the size of the training set increases just modestly, \textbf{showing the validity of our training set generation techniques}.

The importance of a feature in a gradient boosted decision tree model such as LambdaMART can be conveyed by the number of times such feature appears in the internal (non-leaf) nodes of the decision trees that form the tree ensemble. Since our model has 50 trees, each having 15 leaves (and 14 internal nodes), there are 700 branches overall. In Table~\ref{table_frequencies_personal} we present the feature frequency distributions for the trained \emph{w5h-l2r} model. The most frequent feature in our model is the \emph{what} dimension, that represents the content of an object and is scored using \emph{field-based BM25}. Then, features (\emph{who,when}), (\emph{who,how}), and (\emph{who}) appear next, all of them related to \textbf{the \emph{who} dimension, which is expected since personal digital traces are byproduct of actions and events of users (\emph{who}), and are typically focused on user interactions}.

\begin{table}[h!]
\centering
\begin{tabular}{|l||r|}
\hline
Feature & Freq. \\ \hline \hline
what	& 217 \\
who, when	& 91 \\
who, how	& 80 \\
who	& 75 \\
what, how	& 72 \\
what, who, how	& 51 \\
what, when	& 40 \\
what, who	& 28 \\
who, when, how	& 23 \\
when	& 16 \\
what, who, when, 	&  \\
how	& 4 \\
what, when, how	 & 2 \\
how	 & 1 \\
\hline
\end{tabular}
\caption{Feature frequencies for the \emph{w5h-l2r} model and the Personal Data Traces dataset.}
\label{table_frequencies_personal}
\end{table}

The results presented in this section indicates that personal data search can improve greatly by taking into consideration the knowledge the user has about the object being searched. The multidimensional data model, based on the 6 contextual dimensions, proved to be an intuitive and efficient model to unify and link heterogeneous personal digital traces. The advantage of using a learning approach to re-rank search results can be seen by the improvement presented by the \emph{w5h-l2r} approach when compared against both methods, \emph{BM25} and \emph{field-based BM25}. Including a compact feature space based on frequency information resulted in significant improvements. 

\section{Case Studies 2: Enron Email Dataset}

\subsection{Methodology}

\subsubsection{\bf Data Set.}
To verify the validity of our frequency-based learning-to-rank approach over other domains, we have implemented it over an email dataset: the Enron~\cite{enrondataset} dataset. The Enron email dataset contains a total of about 0.5M emails from 158 employees of the Enron Corporation, obtained by the Federal Energy Regulatory Commission after the company collapsed into bankruptcy resulting in a federal investigation.

\subsubsection{\bf Training and Evaluation Query Sets.}
To train and validate our model we use heuristically generated samples as we did in Section~\ref{querysets}. Each query is automatically created by randomly choosing a target object from the evaluation data set. We then choose \emph{d} dimensions, from which we randomly select $v$ random values. For this set of experiments, the training set is comprised by 48000 queries over the Enron dataset and the evaluation set is comprised by 2000 queries. Training and evaluation query sets were built using $v=1$ and 4 different values for parameter $d$: \{\emph{what, who}\}, \{\emph{what, who, when}\}, \{\emph{what, who, when, how}\}, and \{\emph{what, who, how}\}. 

\subsubsection{\bf Evaluation Techniques and Metrics.}
To evaluate the efficacy of the proposed approach for the Enron dataset, we use the same metrics adopted to validate the Personal Digital Data Traces dataset (Section~\ref{subsec:evaltechniques}). The proposed approach is compared against \emph{BM25} and \emph{field-based BM25} and 4 standard evaluation metrics are used: Mean Reciprocal Rank (MRR) of the top-ranked 50 documents, success (precision) of the top 1 retrieved document (success@1), success of the top 3 retrieved document (success@3), and success of the top 10 retrieved document (success@10). Wilcoxon signed-rank test with $p\_value < 0.05$ is used to determine statistically significant differences. 

\subsubsection{\bf Ranking Model.}
To train and evaluate our model, we use the same parameters and steps presented in Section~\ref{subsec:rankingmodel}. We selected the model that shows the best performance on the training set, also taking into account the spread between training and testing metrics. The model selected, that we will call \emph{w5h-l2r}, has the following parameters: number of trees $= 50$; number of leaves $= 15$; minimum leaf support $= 20$; shrinkage $= 0.3$.

\subsection{\bf Results.}
\label{sec:l2r:results:enron}

As with the Personal Digital Data Traces dataset, for the Enron dataset the evaluation set was divided in four different groups by the dimensions in each query: Group 1 $=$ {what, who}; Group 2 $=$ what, who, when; Group 3 $=$ what, who, when, how; Group 4 $=$ {what, who, how}.

\begin{table*}[h!]
\centering
\begin{tabular}{c}
\subfloat[Group 1: what, who]{
\begin{tabular}{|l||cccc|}
\hline
Method           & MRR & s@1 & s@3 & s@10 \\ \hline \hline
BM25             &  0.2591 &	0.1320 & 	0.1120	 & 0.0502   \\ \hline
Field-based BM25 &  0.2549 &	0.1260	 & 0.1053 & 	0.0510 \\
w5h-l2r          &  0.2688 &	0.1560 &	0.1020  &	0.0504   \\ \hline
\end{tabular}
} \\
\subfloat[Group 2: what, who, when]{
\begin{tabular}{|l||cccc|}
\hline
Method           & MRR & s@1 & s@3 & s@10 \\ \hline \hline
BM25             & 0.2346 &	0.1220 &	0.0980 & 	0.0450  \\ \hline
Field-based BM25 & 0.4139 & 	0.2360	  & 0.1767 & 	0.0732  \\
w5h-l2r          & 0.4218 &	0.2495  &	0.1790 & 	0.0727 \\ \hline
\end{tabular}
} \\
\subfloat[Group 3: what, who, when, how]{
\begin{tabular}{|l||cccc|}
\hline
Method           & MRR & s@1 & s@3 & s@10 \\ \hline \hline
BM25             &  0.2422  & 	0.1328	 & 0.0979 & 	0.0449  \\ \hline
Field-based BM25 &  0.4090  &	0.2314 & 	0.1791  &	0.0736  \\
w5h-l2r          & 0.4213  &	0.2575  &	0.1764	  & 0.0744  \\ \hline
\end{tabular}
} \\
\subfloat[Group 4: what, who, how]{
\begin{tabular}{|l||cccc|}
\hline
Method           & MRR & s@1 & s@3 & s@10 \\ \hline \hline
BM25             & 0.2442 & 	0.1060	 & 0.1087	& 0.0478  \\ \hline
Field-based BM25 &  0.2585 &	0.1140	 & 0.1133 & 	0.0492  \\
w5h-l2r          & 0.2477 & 	0.1220  &	0.1020  &	0.0492  \\ \hline 
\end{tabular}
}   \\
\end{tabular}
\caption{MRR, s@1 (success@1), s@3 (success@3),  s@10 (success@10) for groups $1$,$2$,$3$, and $4$ from the Enron dataset. }
\label{table:l2r:groups:enron}
\end{table*}

Table~\ref{table:l2r:groups:enron}a-d, show the MRR, success@1, success@3 and success@10 of each search approach, \emph{BM25} (baseline), \emph{field-based BM25}, and \emph{w5h-l2r}, for Group 1 to 4 of queries. For Group $1$ (Table~\ref{table:l2r:groups:enron}a), the search approach \emph{w5h-l2r} is slight better than \emph{BM25} (baseline) and \emph{field-based BM25} for MRR and success@1. For Group $2$ (Table~\ref{table:l2r:groups:enron}b) and Group $3$ (Table~\ref{table:l2r:groups:enron}c), the search approaches that use the data classified according with our multidimensional data model are considerably more accurate than the keyword-based approach, \emph{BM25}, and the results are statistically significant (Wilcoxon signed-rank test, $p\_value < 0.05$). For those groups, the learned ranking model, \emph{w5h-l2r}, outperforms the \emph{field-based BM25} model for all metrics, the exceptions being success@10 for Group $2$ and success@3 for Group $3$. 
For Group $4$, all approaches had a similar performance. For most scenarios in this group of queries, the features based on the \emph{how} dimension are not contributing to differentiate Enron results.

Table~\ref{table_frequencies_enron} shows the feature frequency distributions for the learned ranking model \emph{w5h-l2r}. The two most frequent features in our model are based on the \emph{what} dimension, that represents the content of an object and is scored using field-based BM25. Then, features related to the \emph{who} dimension appear next, representing the frequency of users (\emph{who}) and the interactions between user/time (\emph{who, when}) and user/topic (\emph{who, what}).

\begin{table}[h!]
\centering
\begin{tabular}{|l||r|}
\hline
Feature & Freq. \\ \hline \hline
what & 273 \\
what, how & 118 \\
who, when & 97 \\
who & 77 \\
what, who & 56 \\
what, when & 25 \\
what, when, who & 19 \\
when & 18 \\
what, when, who, &  \\
how & 15 \\
how & 2 \\
\hline
\end{tabular}
\caption{Feature frequencies for the \emph{w5h-l2r} model and the Enron dataset.}
\label{table_frequencies_enron}
\end{table}

The results discussed in this section show that even though the data and scoring model were proposed with the Personal Digital Data Traces dataset in mind, it can be extended to different domains with promising results.

%
%
\section{Related Work}
\label{relatedwork}

Bell has pioneered the field of life-logging with the project MyLifeBits~\cite{mylifebits,bell-total-recall} for which he has digitally captured all aspects of his life. {\em digi.me}~\cite{digime} is a commercial tool that aims at extending Bell's vision to everyday users. The motivations behind {\em digi.me} are very close to ours; however {\em digi.me} currently only offers a keyword- or navigation-based access to the data; search results can be filtered by service, data type or/and date.

The case for a unified data model for personal information was made in~\cite{haystack,xu03towards}. deskWeb~\cite{deskweb} looks at the social network graph to expand the searched data set to include information available in the social network. Stuff I've Seen~\cite{stuff} indexes all of the information the user has seen, regardless of its location or provenance, and uses the corresponding metadata to improve search results. Our work is related to the wider field of Personal Information Management~\cite{PIMbook}, in particular, search behavior over personal digital traces is likely to mimic that of searching data over personal devices.  Unlike traditional information seeking, which focuses on discovering new information, the goal of search in Personal Information systems is to find information that has been created, received, or seen by the user. 

Email search is a type of personal search that has been well studied. ~\cite{halawi2015} presents a learning-to-rank approach that improves the default ranked-by-time search by taking into consideration time recency and textual similarity to the query. ~\cite{Wang2016} addresses the problem of learning-to-rank from click data in personal search. ~\cite{Zamani2017} explores how to effectively leverage situational contextual features (e.g. time of a search request and the location of the user while submitting the request) to improve personal search quality.  In~\cite{Bendersky2017} the authors leverage user interaction data in a privacy preserving manner for personal search by aggregating non-private query and document attributes across a large number of user interactions. In our scenario, each dataset is comprised by data from only one user, and so it is private by design, not being possible for us to leverage interactions from other users.

In~\cite{Dehghani2017} the authors use classic unsupervised IR models, such as BM25, as a weak supervision signal for training deep neural ranking models. In this context, weak supervision refers to a learning approach that creates it own training data by heuristically retrieving documents for a large query set. Three different neural network-based ranking models are presented, a point-wise ranking model and two pair-wise models. Combinations of neural models with different training objectives and input representations are compared against each other and against the baseline, BM25. The experiments showed that their best performing model significantly outperforms the BM25 model. In our work, we use a similar approach to retrieve matching objects to a given query.

%
%
\section{Conclusions and Future Work}

In this work, we proposed a learning-to-rank approach that uses a compact and efficient frequency-based feature space to rank query results over personal digital traces. The learning model relies on our multidimensional data model to unify and link heterogeneous digital traces using six contextual dimensions: \emph{what, who, when, where, why} and \emph{how}. To overcome the lack of human-labeled training sets, we proposed a combination of known-item query generation techniques and an unsupervised ranking model (\emph{field-based BM25}) to generate our query sets. Experiments over a publicly available email collection and a personal dataset composed by data from a variety of data sources indicates that our frequency-based learning approach can significantly improve the accuracy of search results when compared with a traditional keyword-based approach, \emph{BM25}, and a field-based approach that uses the data parsed according to our multidimensional data model, \emph{field-based BM25}.

Our work shows that using contextual information to model and integrate personal data can lead to more accurate scoring and searching methodologies. By introducing a compact feature set, we made it possible for moderately large datasets to take advantage of modern learning-to-rank techniques that are
usually employed with very large datasets. In addition, we have designed a known-item query generation approach that allowed us to generate query sets on the fly for a
type of application (personal search) in which datasets and training sets are non existent; such approach is private by design, allowing for the end-to-end machine
learning pipeline to run on the device of the user. In the future, the same techniques could be applied to different sets of applications, as we have done with the
Enron dataset.


\bibliographystyle{abbrv}
\bibliography{w5hL2R}

\end{document}